\documentclass[conference,hyphens]{IEEEtran}
\IEEEoverridecommandlockouts
\usepackage{cite}
\usepackage{amsmath,amssymb,amsfonts}
\usepackage{algorithmic}
\usepackage{graphicx}
\usepackage{textcomp}
\usepackage{xcolor}
\usepackage{comment}
\usepackage{threeparttable}
\usepackage{hyperref} 
\usepackage[utf8]{inputenc}
\usepackage{graphicx}
\usepackage{subcaption} 
\usepackage{float}
\usepackage{multirow}
\usepackage{makecell}

\def\BibTeX{{\rm B\kern-.05em{\sc i\kern-.025em b}\kern-.08em
    T\kern-.1667em\lower.7ex\hbox{E}\kern-.125emX}}
\begin{document}

\title{Load-Altering Attacks Against Power Grids: A Case Study Using the GB-36 Bus System Open Dataset}

\author{
    Syed Irtiza Maksud and Subhash Lakshminarayana \\
    School of Engineering, University of Warwick, Coventry, UK \\
     \{syed-irtiza.maksud, subhash.lakshminarayana\}@warwick.ac.uk
    \vspace{- 0.6 cm}
}




\maketitle

\begin{abstract}
The growing digitalization and the rapid adoption of high-powered Internet-of-Things (IoT)-enabled devices (e.g., EV charging stations) have increased the vulnerability of power grids to cyber threats. In particular, the so-called Load Altering Attacks (LAAs) can trigger rapid frequency fluctuations and potentially destabilize the power grid. This paper aims to bridge the gap between academic research and practical application by using open-source datasets released by grid operators. It investigates various LAA scenarios on a real-world transmission network, namely the Great Britain (GB)-36 Zone model released by the UK's National Electricity System Operator (NESO). It evaluates the threshold of LAA severity that the grid can tolerate before triggering cascading effects. Additionally, it explores how Battery Energy Storage Systems (BESS) based fast frequency response services can mitigate or prevent such impacts. Simulations are conducted using DIgSILENT PowerFactory to ensure realistic system representation. The analysis provides several useful insights to grid operators on the LAA impact, such as the influence of the relative locations of BESS and LAA, as well as how delays in attack execution can influence the overall system response. 
\end{abstract}

\section{Introduction}

One of the most pressing concerns in modern Cyber-Physical Power Systems (CPPS) is the increasing use of Internet-of-things (IoT)-controllable high-wattage devices, such as smart heat pumps, Electric Vehicle Charging Stations (EVCS), and home energy management systems etc. While these technologies enhance Demand Flexibility (DF), they also pose significant security risks. Their rapid adoption has outpaced the development of robust security standards, making them vulnerable to cyberattacks and exposing the grid to potential disruptions. The focus of this work is on cyber attacks against these devices at scale, which are referred to as Load-Alerting Attacks (LAAs) \cite{Soltan2018, MalekiLAA2025}. 
Although individual load changes may seem insignificant, a large number of compromised devices can collectively create substantial imbalances in the power system.

One of the earliest investigations into the impact of LAAs on transmission systems was conducted using a simulation-based approach in \cite{Soltan2018}, demonstrating that static LAAs can lead to frequency safety violations. To evaluate the implications on protection mechanisms, \cite{HuangUSENIX2019} performed a realistic assessment focusing on load shedding and N-1 contingency scheduling. Subsequent studies, such as \cite{9726888}, analysed low load and low inertia conditions, revealing that LAAs targeting regions with high load concentrations have the most severe impact on frequency stability. Dynamic LAAs, involving the coordinated switching of load devices synchronized with inter-area oscillatory frequencies, pose a more sophisticated threat. As shown in \cite{10098782}, such attacks can drive the system toward instability. To address both static and dynamic scenarios, \cite{9393479} proposed an analytical framework grounded in second-order dynamic system theory, deriving eigenvalue sensitivities of the frequency control loop and identifying critical nodes for attack deployment. To capture a broader range of potential attack scenarios, \cite{10595409} applied rare event sampling techniques to evaluate the impact of LAAs across various node combinations and load magnitudes.
In \cite{sarieddine2023investigating}  the authors examine the feasibility of exploiting security weaknesses in mobile applications used to operate Electric Vehicle (EV) charging systems. Another study in \cite{10132099} analyzes the execution of Dynamic Load-Altering Attacks (DLAAs) using a distributed Electric Vehicle Service Equipment (EVSE) botnet model, and introduces a time-division framework to assess how communication delays impact the coordination of attack vectors. Several works also focus on the detection and mitigation of LAAs. As this work is primarily concerned with attack impact analysis, we omit these papers for brevity and refer the reader to \cite{MalekiLAA2025} for a detailed review.

The goal of this work is to link theoretical research with real-world use cases through the use of publicly available datasets from grid operators.
Despite the growing body of literature focusing on LAAs, most of these works have been evaluated using synthetic test bus systems (such as the IEEE test systems). While the results offer useful insights, the evaluation of the attack impact on real-world power systems is lacking in research literature. The knowledge is crucial for grid operators and policymakers. For instance, the UK plans to introduce cyber security regulations for organizations that control and operate large-scale loads (such as EV charging firms and flexibility providers) \cite{BEIS_Consult}, for which the knowledge of the threshold load attack that causes unsafe frequency fluctuations is crucial (to determine which organizations must be brought under these regulations).

Furthermore, in recent years, several grid operators worldwide are in the process of introducing BESS-based Fast-Frequency Response (FFR) services to improve the grid's resilience to supply-load imbalances. Examples include BESS-based primary and fast frequency response markets introduced by California Independent System Operator (CAISO), Very Fast Frequency Control Ancillary Services (FCAS) markets introduced by Australia's Australian Energy Market Operator (AEMO), and dynamic containment, moderation and regulation measures introduced by UK's NESO. These FFR services will also improve the resilience of the system to LAAs. To the best of our knowledge, none of the existing works have incorporated the effect of such FFR services in LAA studies.

 To fill the aforementioned gaps, this work performs LAA study employing the GB 36-Zone model, which is an open-source model released by the UK's NESO. The model was developed based on the NESO's 2013 Electricity Ten Year Statement (ETYS) and closely mirrors GB’s actual power system. 
 This paper adopts a simulation approach using DIgSILENT PowerFactory 2024 and utilizes the latest version made available by NESO. Although this paper focuses on the GB power grid, we expect its insights to be valuable to system operators in other regions and to motivate similar studies. The contributions of this paper are as follows:

\begin{itemize}
\item Investigating LAAs on a real-world power system using the GB 36-Zone power system, providing practical insights into how LAAs can disrupt operational stability in actual grid environments.
\item Incorporating realistic BESS-based fast-frequency response services into the model, evaluating their effectiveness in mitigating LAAs and enhancing grid stability during dynamic disturbances. 
\item Determining the critical threshold for system operators of LAA magnitudes that the GB power system can withstand before triggering protective measures.

\end{itemize}
The remainder of this paper is organized as follows: Section II presents the network modeling within the GB context. Section III discusses the framework for LAA impact assessment, including the attack and BESS models. Section IV summarizes the simulation setup, results, and impact analysis. Finally, Section V concludes the paper.

\section{Great Britain’s Power Grid and Frequency Response }

\subsection{GB 36-Zone Power Grid Model and Frequency Control}

The GB-36 open-source DIgSILENT model, released by the UK’s NESO, provides a detailed representation of the power grid across England, Wales, and Scotland  \cite{NesoGB36Model}. The system consists of 36 interconnected zones, with 69 transmission lines operating at 400 kV in a meshed configuration to connect all the zones. Each zone is represented as a subsystem block that includes multiple busbars, all operating at 400 kV. Within each zone, generators, loads, and associated equipment such as transformers, isolators, and circuit breakers are modeled. Each zone also includes a load representing the total demand for that region. The overall GB system in the model consists of 108 busbars and thousands of nodes. There are 76 active synchronous generators and 80 static generators, with several offline generators available for dispatch based on the specific needs of the scenario or case study. The system also includes 8 active interconnectors linking to neighboring countries’ grids. The total system demand is approximately 40 GW, with Zone 8 (near London) having the highest demand at around 3,669.5 MW  \cite{Azizipanah2023GBModel}. The model assumes a constant inertia level (H = 5) for all synchronous generators in the system

The power grid frequency is a key measure of system stability and the operator aims to maintain it close to a nominal value (50 Hz in the context of UK). When the balance between the supply and demand is disrupted, the frequency shifts away from its normal value.  In UK, the operator aims to maintain frequency within ±0.2 Hz of the nominal frequency under normal conditions and within the statutory limit of ±0.5 Hz of nominal frequency. The permissible Rate of Change of Frequency (RoCoF) threshold should not exceed 0.125 Hz/s (or 0.0025 p.u./s for a base frequency of 50 Hz) for more than 500ms. However, through the Accelerated Loss of Mains Change Programme, this limit is being relaxed to 0.5 Hz/s (0.01 p.u./s) and even 1 Hz/s (0.02 p.u./s) in certain locations. Any sudden load change that pushes the frequency beyond these thresholds triggers corrective measures to restore stability. Table \ref{tab:1} outlines the frequency thresholds used in the UK power system and the corresponding control actions.

\begin{table}[h]
    \centering
    \small 
    \caption{Operating Conditions and Actions Based on Frequency Range}
    \label{tab:1}
    \renewcommand{\arraystretch}{1.2}
    \begin{tabular}{|p{1.3cm}|p{3.2cm}|p{3.2cm}|} 
        \hline
        \textbf{Frequency Range (Hz)} & \textbf{Operating Condition} & \textbf{Action Taken} \\
        \hline
        50.00 & Nominal Frequency & No action required \\
        49.8--50.2 & Normal operation & Dynamic supply/demand balancing actions \\
        49.5--50.5 & Statuary limits & Fast \!Frequency Response (FFR) \\
        $\leq$ 48.8 & Under Frequency Load Disconnection & Automatic load shedding \\
        \hline
    \end{tabular}
    \label{tab:freq_action}
\end{table}

\subsection{Battery Energy Storage Systems Operation as Dynamic Frequency Response}
\begin{figure}[htbp]
\centerline{\includegraphics[width=0.75 \linewidth]{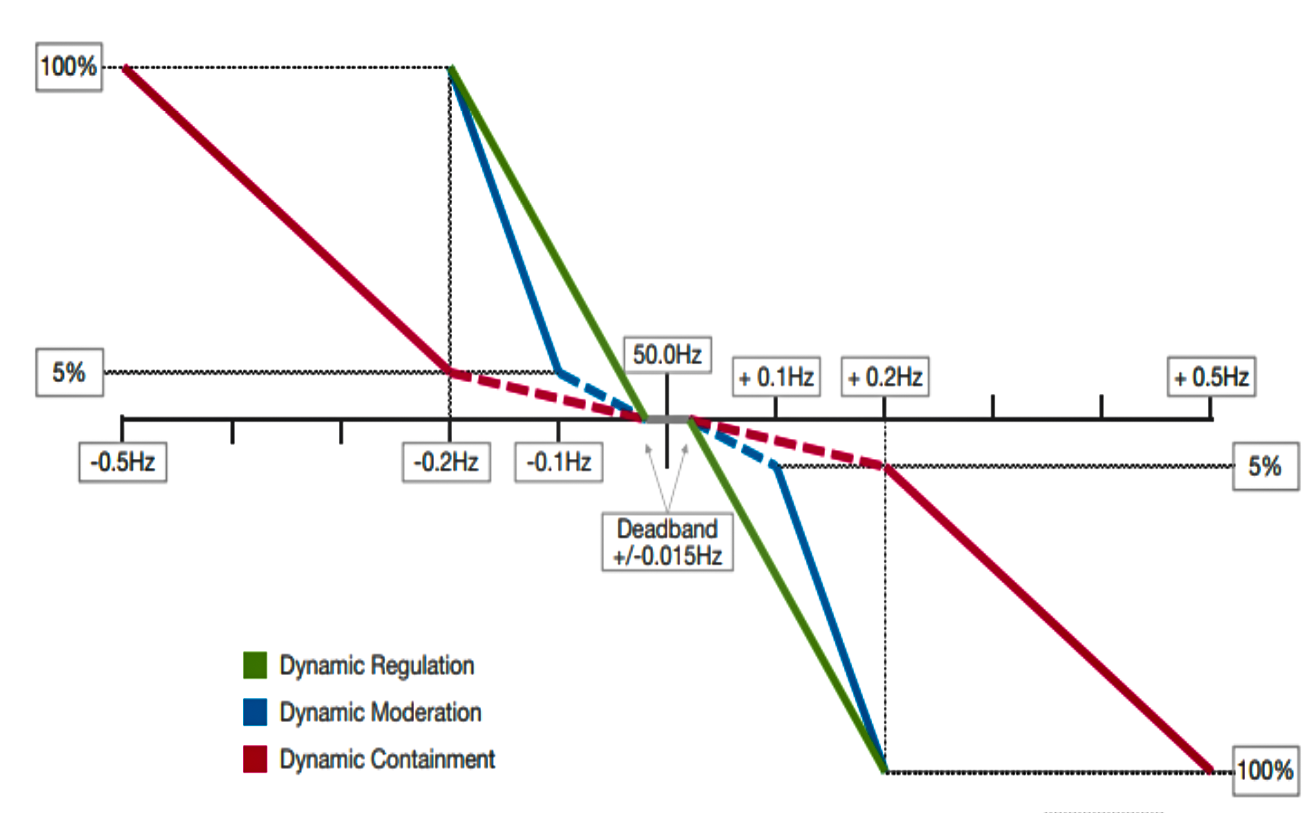}}
\caption{Dynamic frequency response produced in UK \cite{NGDynamicServices}.}
\label{Dynamic frequency}
\end{figure}

With the increasing adoption of renewable energy, maintaining frequency stability after disturbances has become more challenging due to the decline in inertia and spinning reserves. NESO currently aims to maintain a system inertia above 96 GJ (from 140 GJ in 2022). As inertia levels are expected to decline further, there will be an increasing need for artificial inertia or fast energy injections, similar to spinning reserves. BESS have shown promising results in this area, and is the principal form of FFR in the UK. 
They can swiftly absorb and release energy, which is particularly beneficial in low-inertia networks, helping to mitigate oscillations and the RoCoF. To enhance frequency regulation, NESO has introduced three dynamic frequency response mechanisms, all operating within a ±0.015 Hz deadband\cite{NGDynamicServices}.
From Figure \ref{Dynamic frequency}, the operation set points for each product can be understood as follows: Dynamic Containment (DC) is a FFR service designed to provide post-fault support during major frequency deviations for an extended period, activating full power at a ±0.5 Hz deviation. Dynamic Response (DR) and Dynamic Moderation (DM) are both fast-acting services, initiating at 0.5 s with full delivery at 1 s, designed to deliver full power at a ±0.2 Hz deviation from the nominal frequency. DR is a continuously operating service that helps maintain system frequency stability, while DM is a buffer against sudden frequency shifts \cite{NGDynamicServices}.

\section{Methodology}

This section first presents a generic framework outlining the main steps for converting open-source datasets into simulation models compatible with LAA studies. It then delves into the specific LAA case study using the GB-36 bus model. 

\subsection{Framework For LAA Impact Assessment Using Open-Source Datasets}

\begin{figure}[!t]
\centerline{\includegraphics[width=1\linewidth]{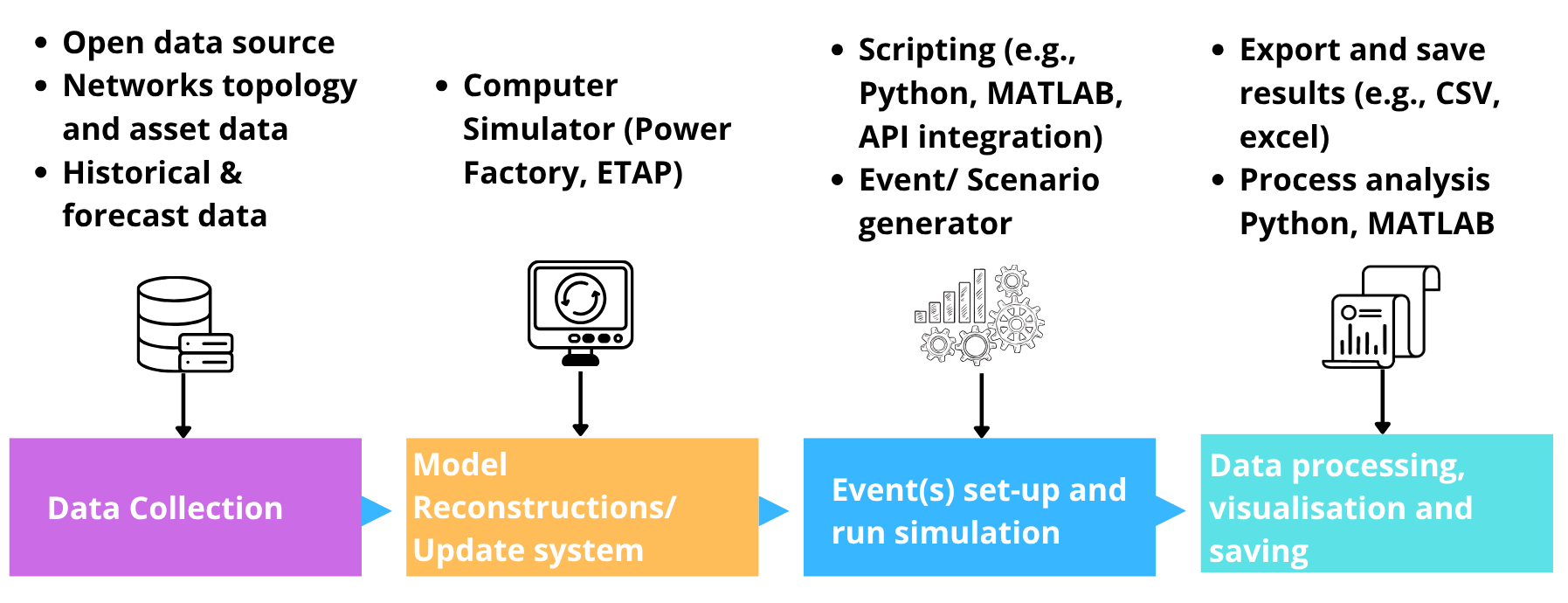}}
\caption{Process flow for modeling and evaluating LAAs.
\vspace{- 5mm}
}
\label{Process Flow}
\end{figure}

 Figure \ref{Process Flow} presents an overview of the four stages involved in using open-source datasets of LAA studies. The first stage involves collecting network and operational data from open-source repositories, including grid topology, generation, and load profiles. While the focus of this work is on the GB-36 Zone model from the NESO, various Distribution System Operators (DSOs), like the UK Power Networks (UKPN) and Electricity Northwest (ENW) also offer complementary datasets for more localized grid information under their ``Open data portals'' \cite{ENWL_DataPortal}, \cite{UKPN_OpenData}.

 However, the datasets provided are often not directly compatible with existing power system simulation platforms. This challenge is addressed in the second stage, which focuses on updating or reconstructing the network model. When detailed topology or system parameters are missing, manual reconstruction is required using available data and engineering judgment. For instance, UK DSOs typically do not provide the actual grid model in tools like DIgSILENT in their open-source portal. However, open-access resources such as long term development statements and the embedded generation capacity register offer valuable information on grid topology and connected load and generation data, which can be used to build simulation models. Additionally, data on substation locations and demand locations help identify critical network points. Curtailment records and forecast data provide insights into historical load patterns and constraints that can support realistic scenario modelling.
 In this study, the GB 36-zone model was publicly available from NESO, but it was modified and updated as needed to meet the specific requirements of the study.
 
In the third stage, various LAA scenarios are created by altering load distributions across zones to observe the system’s response to coordinated disturbances and identify critical thresholds. While all simulations in this work were performed manually to analyze cascading failures, this stage can be automated using scripting tools (e.g., Python or MATLAB) to support randomized sampling, adversarial case generation, and broader scenario coverage. This is especially important, given that power grids are naturally resilient to large-scale load changes, given the N-1 design criteria. Thus, uncovering impactful LAA requires advanced sampling techniques \cite{10595409}. 

The final stage involves processing and storing simulation results for visualization and analysis. Outputs can be archived for post-event review and illustrated through visual dashboards to highlight critical dynamics. Automating this pipeline enhances repeatability, reduces manual effort, and supports integration with data-driven methods for predicting system vulnerabilities.

\subsection{LAA Study Using the GB-36 Model}

\begin{figure}[!t]
\centerline{\includegraphics[width=0.8\linewidth]{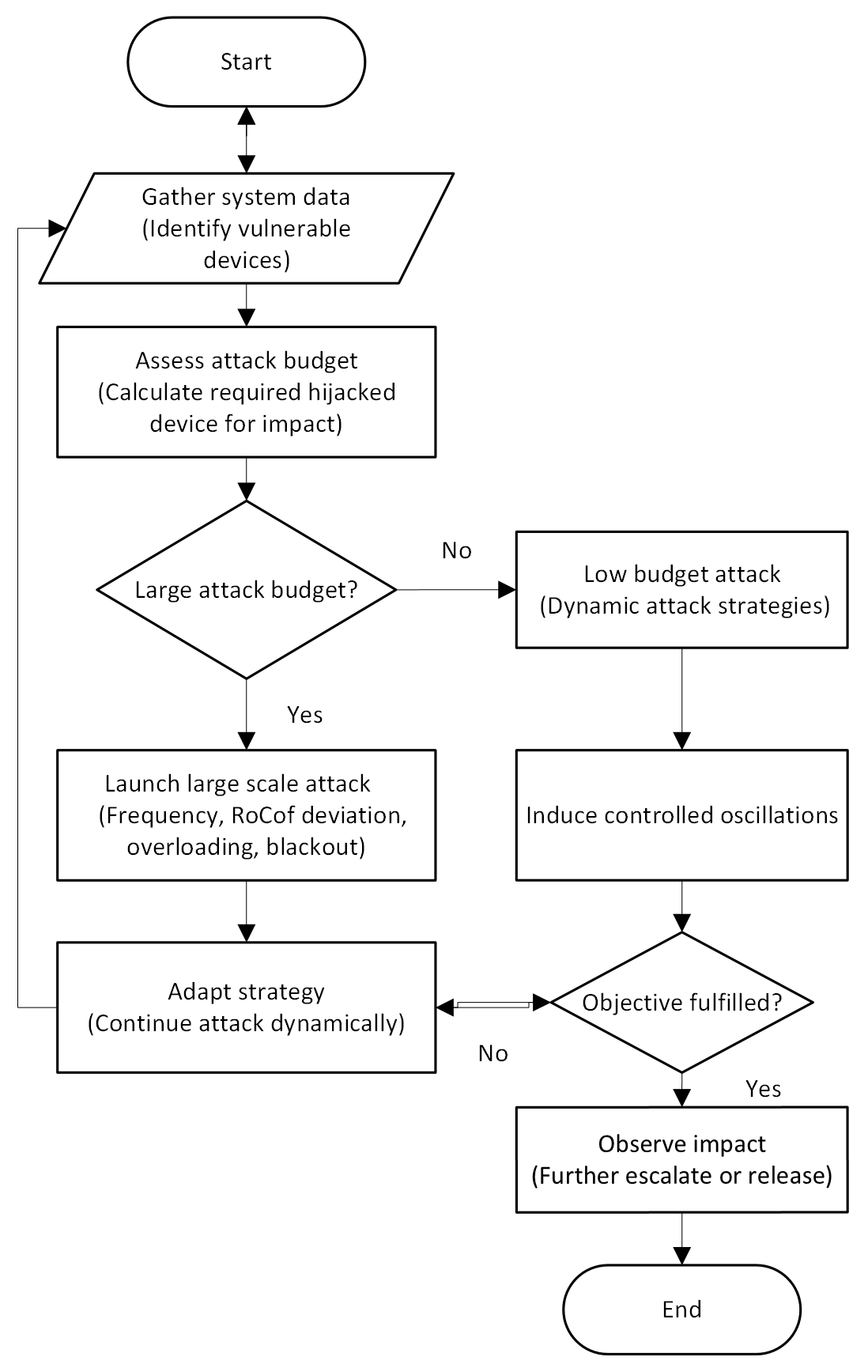}}
\caption{Decision Flowchart for an Adversary’s LAA. }
\label{flowchart}
\vspace{- 6 mm}
\end{figure}

We consider a scenario where an attacker has the ability to modify controllable loads connected to the load buses and
can access their frequency measurements. 
Note that throughout the paper, we assume that the attacker has already compromised the load and executes the LAA, and omit the details of the threat modeling (details of which can be found in \cite{MalekiLAA2025}).
Some loads may be protected or beyond the attacker's reach, making them unalterable. These are excluded from consideration, while the focus is on compromised loads with weak controls that fall within the attacker's budget. Figure \ref{flowchart} illustrates possible attack operation which begins with identifying vulnerable devices and selecting either large-scale or low-budget dynamic strategies depending on available resources. The attack evolves based on system feedback and continues until the desired impact is achieved.

The GB 36-zone model developed by NESO does not have BESS integrated into the network by default. To evaluate the frequency response capability of BESS, we incorporated BESS units at selected locations using a pre-configured model from the DIgSILENT PowerFactory library, listed under Storage Systems as “DIgSILENT BESS FrequencyCtrl 10kV 30MVA.” This model is implemented using the \texttt{ElmGenstat} object class, which represents a static generator. Although \texttt{ElmGenstat} itself is a static generator, the DIgSILENT BESS model includes embedded control blocks that enable frequency-responsive active power support. For the purpose of our simulations, we configured each BESS unit’s power rating and droop control parameters in accordance with standard frequency response requirements (DR and DC type). Each unit was manually connected to the transmission network through a 400/10 kV step-down transformer to ensure proper voltage level alignment and seamless integration into the grid. Further details on the BESS model can be found in \cite{digsilent_bess}. 

In our simulations, five BESS units are allocated for DR service at Zones 1, 8, 20, 25A, and 27W, while five units are deployed for DC service at Zones 3, 8, 9, 15, and 25. The BESS ratings are adjusted individually for each case analysis. As the deadband and full activation thresholds of DR and DM mode are nearly identical, DM is excluded from this study to maintain a focused and concise analysis. For this analysis, we consider a static load increase type LAA, as this type of attack is the most impact full. The attack was initiated at Zone 8 at 1 second. The highest RoCoF value, and system frequency observed across the system was recorded.

\section{Case Study Results}


\begin{table*}
\centering
\caption{Threshold LAA magnitude to cause unsafe frequency excursion for the GB-36 bus system with and without BESS. In each case, we consider $500$ MW BESS either in DC or DM mode.}
\label{tab:2}
\begin{tabular}{|c|c|c|c|c|c|c|c|}
\hline
\parbox[c]{2.7cm}{\centering \textbf{Threshold}\\\textbf{Analysis}} &
\parbox[c]{2.2cm}{\centering \textbf{Frequency}\\\textbf{Limit (Hz)}} &
\parbox[c]{2.5cm}{\centering \textbf{Minimum}\\\textbf{LAA Level (MW)}} &
\parbox[c]{2cm}{\centering \textbf{RoCoF}\\\textbf{p.u./s}} &
\parbox[c]{2cm}{\centering \textbf{Frequency}\\\textbf{(Nadir (Hz))}} &
\parbox[c]{2cm}{\centering \textbf{Nadir}  \\\textbf{Time (s)}} &
\parbox[c]{2cm}{\centering \textbf{Settling}\\\textbf{Freq. (Hz)}} \\
\hline
\textbf{Without BESS} & 49.80 & 660.00     & -0.1227 & 49.79 & 5.98   & 49.91 \\
                      & 49.50 & 900.00     & -0.1658 & 49.49 & 12.89  & 49.88 \\
                      & 48.80 & 1027.46 & -0.1950 & 48.80 & 108.00 & 49.89\textsuperscript{*} \\
\hline
\textbf{With DC}      & 49.80 & 679.00     & -0.0882 & 49.79 & 5.92   & 49.92 \\
                      & 49.50 & 972.42  & -0.1347 & 49.47 & 10.25  & 49.85 \\
                      & 48.80 & 1541.20  & -0.2160 & 48.80 & 21.92  & 49.37\textsuperscript{*} \\
\hline
\textbf{With DR}      & 49.80 & 1027.46 & -0.1349 & 49.79 & 4.63   & 49.89 \\
                      & 49.50 & 1376.00    & -0.1837 & 49.47 & 49.47  & 49.83 \\
                      & 48.80 & 1541.20  & -0.2075 & 48.80 & 45.00  & 49.84\textsuperscript{*} \\
\hline
\end{tabular}
\begin{tablenotes}
\small
\item \textsuperscript{*} Final settling frequency after UFLS (5\% load shedding).
\end{tablenotes}
\end{table*}

\begin{table*}
\centering
\caption{System Response to Load-Altering Attacks with Different BESS Sizes.}
\label{tab:3}
\begin{tabular}{|>{\centering\arraybackslash}p{2.7cm}|>{\centering\arraybackslash}p{2.2cm}|>{\centering\arraybackslash}p{2.5cm}|>{\centering\arraybackslash}p{2.0cm}|>{\centering\arraybackslash}p{2.0cm}|>{\centering\arraybackslash}p{2.0cm}|>{\centering\arraybackslash}p{2.0cm}|}
\hline
\makecell{\textbf{BESS Size,} \\ \textbf{Equal DC \& DR Mix}} & \makecell{\textbf{Frequency} \\ \textbf{Limit (Hz)}} & \makecell{\textbf{Minimum} \\ \textbf{LAA Level (MW)}} & \makecell{\textbf{RoCoF} \\ \textbf{(P.U./s)}} & \makecell{\textbf{Frequency} \\ \textbf{Nadir (Hz)}} & \makecell{\textbf{Nadir} \\ \textbf{Time (s)}} & \makecell{\textbf{Settling} \\ \textbf{Freq. (Hz)}} \\
\hline
\multirow{3}{*}{400MW} & 49.80 & 807.29   & -0.1076 & 49.79 & 4.54  & 49.91 \\
                       & 49.50 & 1115.52 & -0.1509 & 49.49 & 11.22 & 49.85 \\
                       & 48.80 & 1431.10   & -0.1969 & 48.80  & 44.54 & 49.49\textsuperscript{*}  \\
\hline
\multirow{3}{*}{500MW} & 49.80 & 880.68 & -0.1169 & 49.79 & 4.044 & 49.90 \\
                       & 49.50 & 1174.24 & -0.1582 & 49.48 & 11.44 & 49.82 \\
                       & 48.80 & 1541.20   & -0.2116 & 48.80  & 28.41 & 49.44\textsuperscript{*}  \\
\hline
\multirow{3}{*}{600MW} & 49.80 & 910.03 & -0.1011  & 49.79 & 4.324 & 49.90 \\
                       & 49.50 & 1221.94  & -0.1637 & 49.49 & 11.28 & 49.83 \\
                       & 48.80 & 1632.92 & -0.2235 & 48.80  & 32.88 & 49.92\textsuperscript{*} \\
\hline
\end{tabular}
\begin{tablenotes}
\small
\item\textsuperscript{*}Final Settling Frequency after UFLS scheme (5\% Load shedding).
\end{tablenotes}

\end{table*}

\begin{table*}
\centering
\caption{Influence of LAA Location on Frequency Stability During LAA Event.}
\label{tab:4}
\begin{tabular}{|>{\centering\arraybackslash}p{2.7cm}|>{\centering\arraybackslash}p{2.2cm}|>{\centering\arraybackslash}p{2.5cm}|>{\centering\arraybackslash}p{2.0cm}|>{\centering\arraybackslash}p{2.0cm}|>{\centering\arraybackslash}p{2.0cm}|>{\centering\arraybackslash}p{2.0cm}|}
\hline
\makecell{\textbf{BESS Size,} \\ \textbf{Equal DC \& DR Mix}} & \makecell{\textbf{LAA Activation} \\ \textbf{Location}} & \makecell{\textbf{LAA Level (MW)}} & \makecell{\textbf{RoCoF} \\ \textbf{(P.U./s)}} & \makecell{\textbf{Frequency} \\ \textbf{Nadir (Hz)}} & \makecell{\textbf{Nadir} \\ \textbf{Time (s)}} & \makecell{\textbf{Settling} \\ \textbf{Freq. (Hz)}} \\
\hline
\multirow{5}{*}{500 MW} &  Zone 8  & 880.68 & -0.1169 & 49.79 & 4.044 & 49.90 \\
                       &   Zone 1  & 880.68 & -0.1098 & 49.78 & 4.942 & 49.90 \\
                       &   Zone 15  & 880.68   & -0.0668 & 49.78  & 4.984 & 49.90 \\
                        &  Zone 20  & 880.68 & -0.0462 & 49.78 & 4.887 & 49.90 \\
                       &   Zone 27W  & 880.68   & -0.1844 & 49.79  & 4.816 & 49.90 \\

\hline
\end{tabular}
\end{table*}

\paragraph{Stability Limits Under LAAs}

To ensure secure operation, system operators require insight into the levels of LAA that can drive the system beyond frequency safety limits. Table \ref{tab:2} outlines these critical LAA thresholds which breach the safety margins. The values presented in the table result from iterative testing of different LAA magnitudes applied to the system, focusing on Zone 8. For both DR and DC modes, the BESS capacity is considered as 500MW. In the absence of any BESS, the system exhibits the poorest frequency performance, with low nadirs, higher RoCoF, and a delayed settling period, indicating a greater risk of instability. This is particularly evident at lower frequency thresholds, such as 48.8 Hz, where UFLS (Under-Frequency Load Shedding) is triggered. 
When BESS is integrated, at the same LAA levels, DR  mode consistently shows superior performance over DC mode, achieving higher nadirs and lower RoCoF, thereby reducing the risk of frequency collapse and enabling faster stabilization. Interestingly, even when the LAA percentage remains the same, the type of BESS mode leads to noticeably different frequency dynamics, emphasizing the importance of selecting the appropriate control strategy. It is also observed that while a 1027.46MW of LAA without BESS causes a critical limit violation (48.8 Hz), which is the only violation of the 49.8Hz limit for DR mode. This analysis highlights the critical role of BESS not just in mitigating frequency deviations but also in enhancing the system's threshold for tolerating disturbances, provided the control mode is properly optimized. Table III further expands this analysis by presenting different threshold levels for varying BESS sizes, keeping a balanced 50-50 mix of DR and DC dynamic services. The values show marginal variation when LAA is applied from other zones, as will be demonstrated in a subsequent section. However, this threshold serves as a good indicator of the overall system’s LAA sensitivity.

\begin{table*}
\centering
\caption{System Response to Load-Altering Attacks with BESS Activation in the  LAA Zone.}
\label{tab:5}
\begin{tabular}{|>{\centering\arraybackslash}p{2.7cm}|>{\centering\arraybackslash}p{2.2cm}|>{\centering\arraybackslash}p{2.5cm}|>{\centering\arraybackslash}p{2.0cm}|>{\centering\arraybackslash}p{2.0cm}|>{\centering\arraybackslash}p{2.0cm}|>{\centering\arraybackslash}p{2.0cm}|}
\hline
\makecell{\textbf{BESS Size,} \\ \textbf{Equal DC \& DR Mix}} & \makecell{\textbf{BESS Activation} \\ \textbf{Location}} & \makecell{\textbf{LAA Level (MW)}} & \makecell{\textbf{RoCoF} \\ \textbf{(P.U./s)}} & \makecell{\textbf{Frequency} \\ \textbf{Nadir (Hz)}} & \makecell{\textbf{Nadir} \\ \textbf{Time (s)}} & \makecell{\textbf{Settling} \\ \textbf{Freq. (Hz)}} \\
\hline
\multirow{3}{*}{500 MW} & \multirow{3}{*}{Zone 8} & 880.68 & -0.0981 & 49.82 & 3.944 & 49.91 \\
                       &         & 1174.24 & -0.1381 & 49.63 & 7.514 & 49.86 \\
                       &  & 1541.20   & -0.1844 & 48.80  & 39.57 & 49.73\textsuperscript{*}  \\
\hline
\end{tabular}
\begin{tablenotes}
\small
\item\textsuperscript{*}Final Settling Frequency after UFLS scheme (5\% Load shedding).
\end{tablenotes}

\end{table*}
\paragraph{Influence of LAA Location on Frequency Stability During LAA Event}

This subsection investigates how the location of an LAA affects system frequency response when the total attack magnitude and BESS configuration remain constant. A fixed LAA level of 880.68 MW is applied at different zones within the GB-36 system. Table \ref{tab:4} presents the system response for LAAs originating in Zones 1, 8, 15, 20, and 27W. RoCoF spans from -0.0462 per unit per second in Zone 20 to -0.1844 per unit per second in Zone 27W. This variation illustrates that certain zones induce more abrupt frequency declines than others, even though the disturbance magnitude remains identical. Frequency nadirs also vary slightly, with values ranging between 49.78 Hz and 49.79 Hz, suggesting minor differences in the severity of the initial frequency drop. The difference is likely influenced by local factors such as load concentration, inertia distribution, and electrical proximity to both generation and BESS support. Although all scenarios recover to the same settling frequency, the transient dynamics could have significant implications for protection mechanisms, especially those relying on RoCoF and nadir-based thresholds. These results emphasize the importance of spatial considerations when assessing system vulnerability to coordinated load-based attacks. Mitigation strategies that rely solely on total disturbance magnitude may overlook the localized impact of LAAs.

\begin{figure} [t!]
\centerline{\includegraphics[width=\linewidth]{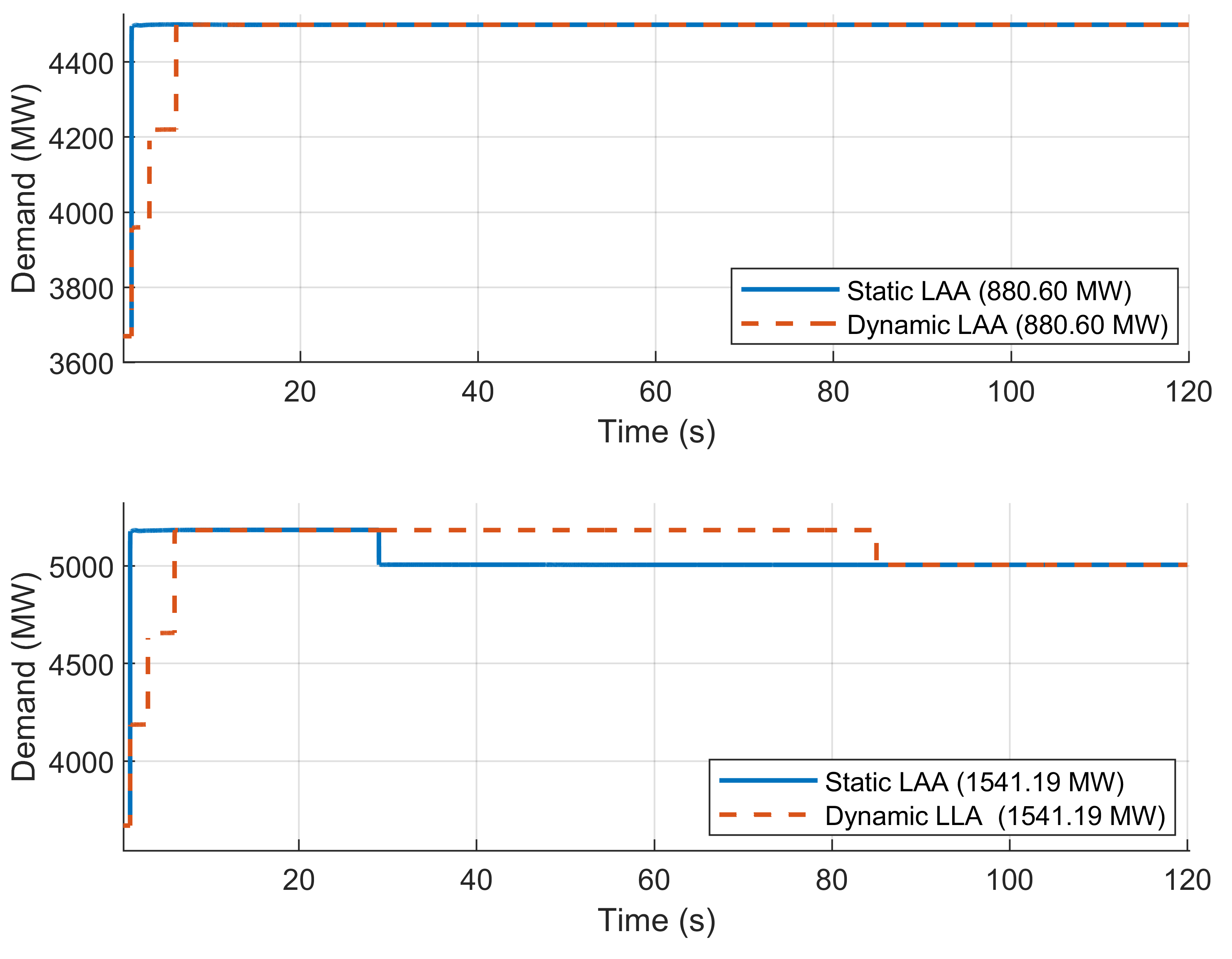}}
\caption{Load Injection Profile under Static and Dynamic LAA.}
\label{3}
\vspace{-0.3 cm}
\end{figure}

\begin{figure} [t!]
\centerline{\includegraphics[width=\linewidth]{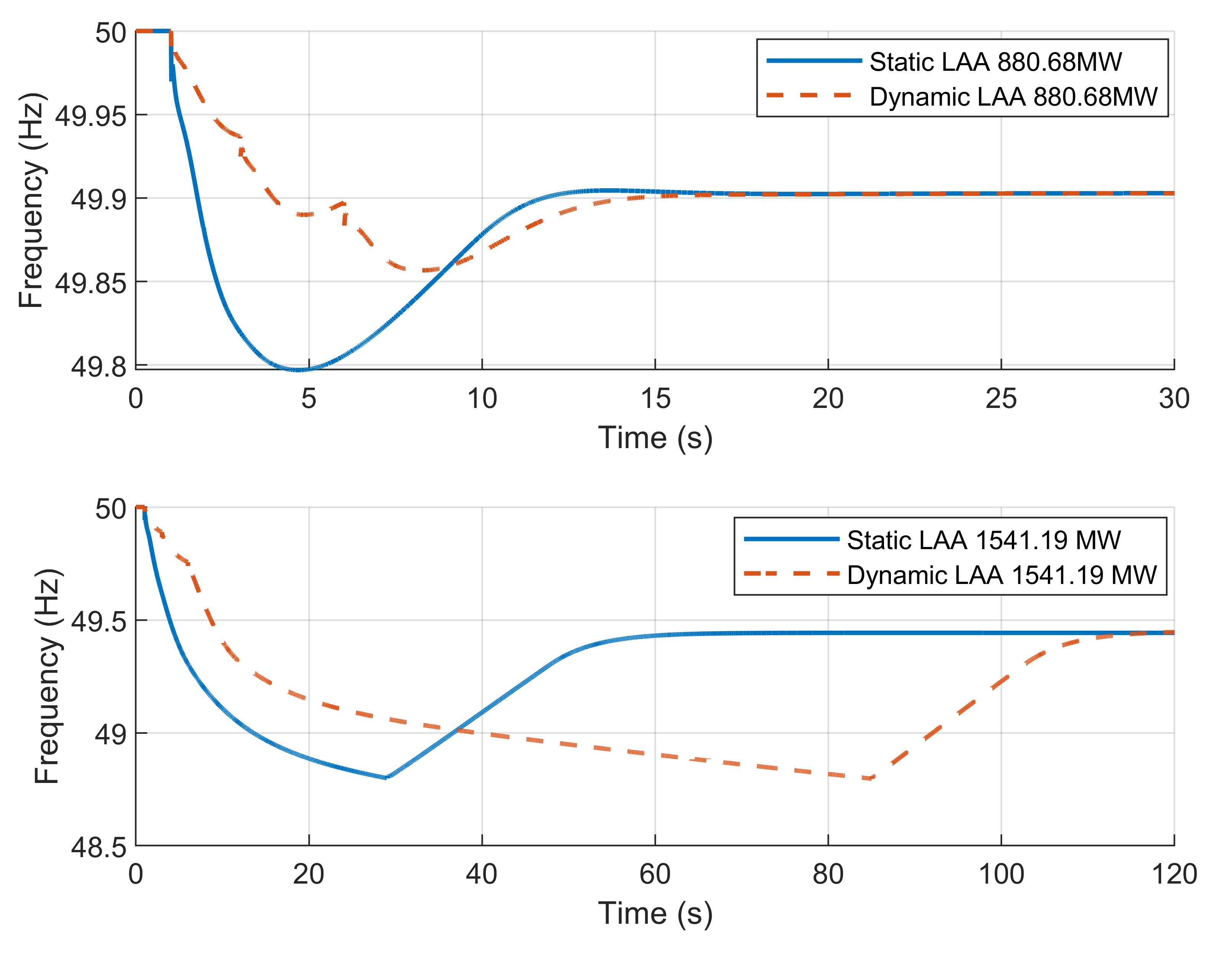}}
\caption{ Frequency Response Comparison under Static and Dynamic LAA.}
\label{4}
\vspace{-0.4 cm}
\end{figure}

\paragraph{Influence of BESS Activation Zone on Frequency Stability During LAA}

When BESS is activated within the same LAA zone as the LAA (Zone 8), the system shows a more stable frequency response and a lower RoCoF. A comparison between \ref{tab:3} and \ref{tab:5} shows that even with the same LAA level, identical BESS size, and similar DC and DR mix (250MW in DR and 250MW in DC modes), the system response differs based on the BESS activation zone. This variation is particularly evident at the highest LAA level of 1541.20 MW, where the frequency drops to the critical threshold of 48.80 Hz at 39.57 seconds, which is a significantly longer delay. After UFLS, the frequency settles at 49.44 Hz compared to 49.73 Hz when BESS is activated within the same zone. These findings highlight the importance of proper BESS placement in mitigating the impact of LAAs. Based on these insights, BESS location can be further analysed for its role in overall system stability.

\paragraph{Impact of Static vs. Dynamic LAAs on System Frequency Response}
We compare the effects of static and dynamic LAA using two critical load levels: 880.68 MW and 1541.20 MW. As shown in Table \ref{tab:3}, these levels cause the system frequency to drop below 49.8 Hz and 48.8 Hz, respectively, even with a 500 MW BESS composed of an equal mix of DR and DC. Figure \ref{3} and \ref{4} show that when the LAA is applied as a static step, the frequency drops sharply due to the system's inability to respond in time. In contrast, the dynamic LAA applies the same total load change in three steps at 1, 3, and 6 seconds, allowing system inertia, spinning reserves, and the BESS to respond more effectively. For the 880.68 MW dynamic LAA, the frequency remains within acceptable limits (above 49.8 Hz). For the larger 1541.20 MW case, the dynamic attack results in breaching the 48.8 Hz limit at 85 seconds, compared to just 28.41 seconds in the static case. With a 5-second interval between the first and last load steps in the dynamic attack, this approach provides a significant time margin of approximately 56 seconds for the system to initiate defensive actions.  These results suggest that incorporating built-in response delays in smart devices can significantly reduce the impact of sudden load-altering attacks. An interesting observation from this analysis is that, for both static and dynamic types of attacks, the final settling frequency remains the same for the exact LAA level.

\section{Conclusion}

This paper investigated the impact of LAAs on a real-world transmission network, specifically within the UK context, using the open-source GB 36-Zone network model in DIgSILENT PowerFactory. The analysis showed how varying attack magnitudes can significantly affect grid stability, causing rapid frequency deviations and potentially triggering cascading failures. The results demonstrated the network’s ability to handle large sudden imbalances from different attack scenarios and identified a critical LAA severity threshold beyond which stability is compromised. The study also highlighted that BESS can help mitigate or prevent these effects, depending on placement and size. Delays in attack execution were found to influence system response, underlining the dynamic nature of these threats. Threshold limits for cascading failure scenarios were identified by manually inputting values across multiple simulations until the critical point was reached. Future work includes developing an automated LAA event script to reduce manual processing and combining the physical impact of attacks with cyber threat and risk scores for a comprehensive risk analysis.

\section{Acknowledgment}
The work was supported by the University of Warwick's Policy Support Grant.

\vspace{-2 mm}
\bibliographystyle{IEEEtran}
\bibliography{IEEEabrv,bibliography}

\end{document}